\RequirePackage{fix-cm}
\documentclass[amsmath,amssymb]{natureprintstyle}

\usepackage{amsmath} 
\usepackage{amsthm} 
\usepackage{graphicx}
\usepackage{subcaption}
\usepackage{appendix}
\usepackage{color}

\usepackage{tikz,pgfplots}

\title{3-D Printed Swimming Microtori for Cargo Transport and Flow Manipulation}

\author{Remmi Baker,$^{1,\dagger}$ Thomas Montenegro-Johnson,$^{2,\dagger}$ Anton D. Sediako,$^3$ Murray J. Thomson,$^3$ Ayusman Sen,$^4$ Eric Lauga$^5$\& Igor. S. Aranson$^{* 4,6,7}$}

\begin{document}

\maketitle

\begin{affiliations}
 \item Department of Material Science and Engineering, The Pennsylvania State University, University Park, PA 16802, USA
 \item School of Mathematics, University of Birmingham, Birmingham B15 2TT, UK
 \item Department of Mechanical Engineering, University of Toronto, Toronto M4Y 2H9, CA
 \item Department of Chemistry, The Pennsylvania State University, University Park, PA 16802, USA
  \item Department of Applied Mathematics, University of Cambridge, Cambridge CB2 1TN, UK
  \item Department of Mathematics, The Pennsylvania State University, University Park, PA 16802, USA
 \item Department of Biomedical Engineering, The Pennsylvania State University, University Park, PA 16802, USA
\end{affiliations}

\begin{abstract}
Through billions of years of evolution, microorganisms mastered unique swimming behaviors to thrive in complex fluid environments. Limitations in nanofabrication have thus far hindered the ability to design and program synthetic swimmers with the same abilities. Here we encode multi-behavioral responses in artificial swimmers such as microscopic, self-propelled tori using nanoscale 3D printing. We show experimentally and theoretically that the tori continuously transition between two primary swimming  modes in response to a magnetic field. The tori also manipulate and transport other artificial swimmers, bimetallic nanorods, as well as passive colloidal particles. In the first behavioral mode, the tori accumulate and transport nanorods; in the second mode, nanorods align along the tori’s self-generated streamlines. Our results indicate that such shape-programmed microswimmers have the potential to manipulate biological active matter, e.g. bacteria or cells.
\end{abstract}

\section{Introduction}
Active systems consist of self-propelled agents, e.g. motile organisms or synthetic particles, that convert energy into mechanical  movement. The intrinsic out-of-equilibrium nature of active systems leads to complex behaviors which often cannot be captured by thermodynamic description. Examples of active matter include  bacterial suspensions, \cite{dombrowski2004self,sokolov2007concentration} starling murmurations, \cite{cavagna2010scale} and fish schooling. \cite{parrish2002self}

Investigation of active matter systems in the limit of low Reynolds number, i.e. microscopic systems, has shown growing interest. \cite{lauga2009hydrodynamics} Microorganisms have been demonstrated to readily adapt to environmental changes. In particular, bacterial systems show a wide variety of complex behaviors, including spontaneous alignment in the presence of chemical gradients \cite{brenner1998physical} and altering rheological properties of the fluid. \cite{sokolov2009reduction, sokolov2009enhanced, lopez2015turning} Translating these complex behaviors to artificial systems is especially attractive for applications in fluid transport, small-scale mixing, and targeted cargo delivery  \cite{patra2013intelligent, sundararajan2008catalytic} but is hard due to intrinsic nanofabrications limitations. 

Recent advances in self-propelled swimming particles have shown micro- and nanomotors are capable of biomimicking bacterial systems. For instance, chemically-powered bimetallic nanorods randomly swim in their environments, in a manner somewhat akin to the swimming of microorganisms like E. coli. \cite{Wang2013, Wang2013b, Paxton2006, wang2006bipolar, wang2013small} In addition, the micromotors (artificial swimmers) autonomously reorient to swim against flows, replicating biological rheotaxis behavior.\cite{Ren2017a, palacci2015artificial, simmchen2016topographical, Zheng2017, Uspal2015} However, these "simple'' artificial swimmers lack the swimming fidelity and multi-responsive behaviors of their biological counterparts; leaving much to be realized before using the micromotors for applications. \cite{Wang2013, Suematsu2018, Moran2011, Ebbens2013, Sabrina2018, Brooks2018, gao2010magnetically, Wang2018a, gao2013bioinspired, li2014template} 

Here,  we 3D print via 2-photon lithograthy precise chemically-powered Janus (anisotropically patterned)  platinum microtori with multi-responsive behaviors. The design is inspired by the "autophoretic torus'' \cite{schmieding2017autophoretic}; in turn by the "smoking ring'' propulsion proposed by Purcell \cite{purcell1977life} and later theoretically demonstrated by Leschansky et al. \cite{leshansky2008surface}  We find that microtori will spontaneously break symmetry to swim linearly across a surface at a constant velocity--radically different from the random swimming and tumbling behaviors seen in other micromotors, such as bimetallic nanorods or Janus spheres. 

We also demonstrate precise control for active cargo collection-transport and active swimmer manipulation by utilizing shape-we encode responses and material properties to program specific behaviors in the tori. Specifically, we harness the magnetific properties presence in our nickel binding layer to tune the tori's axial orientation and as a result alter the particles' behavior. This simple external control of the orientation (and by extension the behavior) of the tori gives rise to two distinct modes: the first mode is a horizontally-oriented and linearly translating tori and in the second mode the tori are vertically-oriented and can be directed linearly or cycloidally. We then experimentally demonstate the unique behaviors of each mode to manipulate other active particles, bimetallic and self-propelled nanorods. In the first mode the horizontally-orieinted tori actively collect and transport bimetallic nanorods to specified locations. In the second mode, the vertically-orieinted tori manipulate the swimming behavior of the bimetallic nanorods by aligning nearby nanorods along the tori's streamlines.

To rationalize experimental observations, we develop a purely hydrodynamic and propulsion mechanism independent model to account for the new emergent phenomena in swimming microtori near boundaries. We determine two main mechanisms responsible for the swimming behavior: self-induced slip velocities across the surface and electrostatic potentials. The induced slip-velocities provides the necessary force for the tori to hover above a surface. Then the electrostatic potential across the anode-cathode over a charged surface tilts the tori. The combined slip velocities at a tilt then leads to directed, linear swimming. 

Our experimental results demonstrate a plethora of new dynamic behaviors that arose in precisely 3D printed and chemically-powered autonomous microswimmers. We also develop a purely hydrodynamic model to predict new shapes and site-specific surface functionalizations to achieve a set of specific swimming and multi-responsive behaviors in 3D printed artificial swimmers. Thus, our experimental and modelling insights can be applied to synthetic microswimmers powered by alternative mechanisms. In particular, our findings can be applied to biological systems by utilizing bio-compatible propulsion mechanisms, such as mounted enzymes \cite{zhao2018powering} or light \cite{palagi2016structured}. These biocompatible, 3D printed microswimmers would then be able to interface with manipulate biological active matter, e.g. cells or bacteria--leading to the development of smart and adaptive cell transport and therapy.

\begin{figure}
	\includegraphics[width=0.6\linewidth]{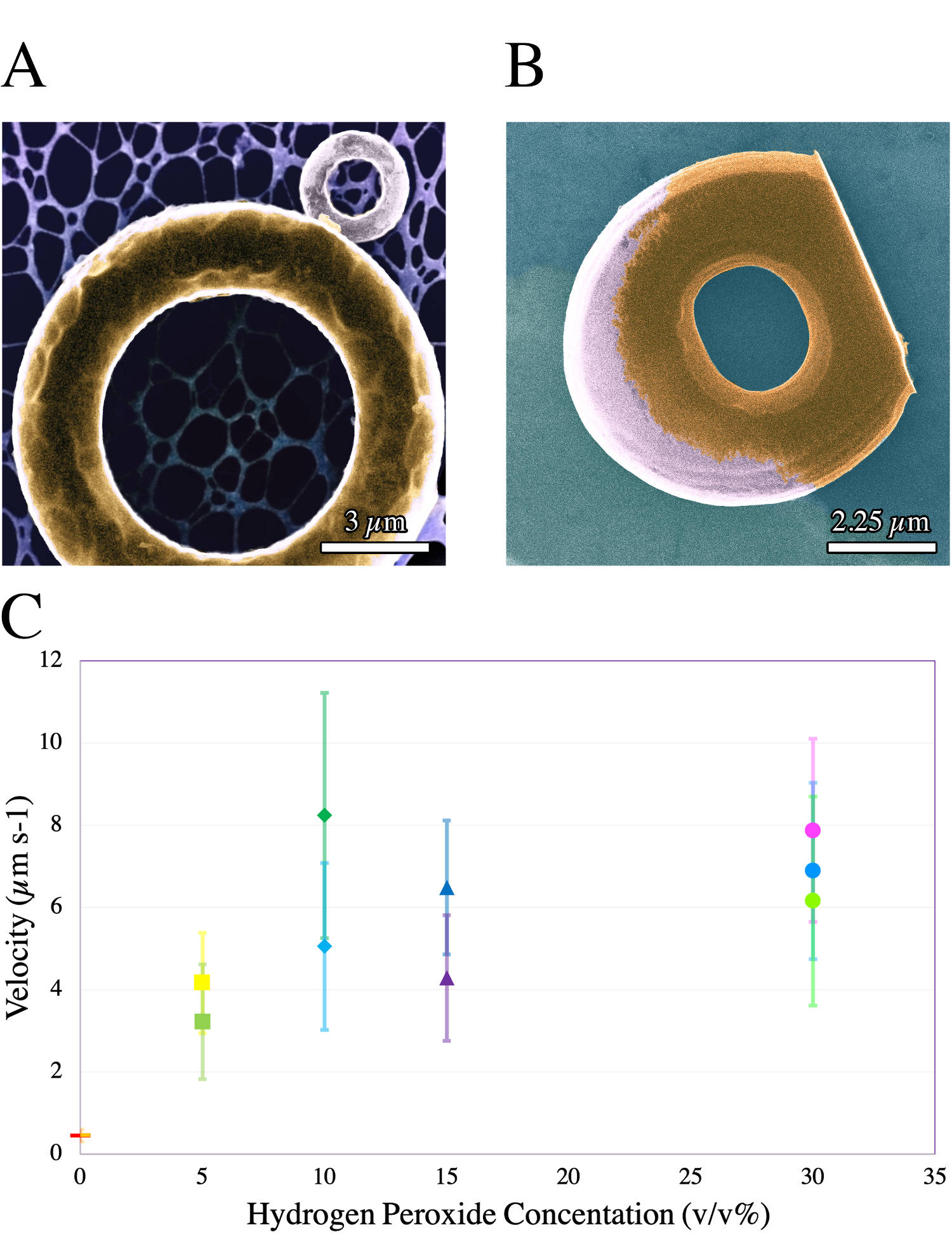}
	\caption{{\bf Tori with two types of surface coating.} (A) A High-resolution transmission electron microscope (HRTEM) image of ``glazed'' Janus tori on a carbon lacey TEM grid. The metallic cap has been applied to the top of the tori. The scale bar is 3 $\mu$m. (B) HRTEM image of a ``dipped'' Janus tori on a carbon TEM grid. The bottom of the tori is ``cut off'' at the bottom to provide a stable base during printing and the metal evporation. The scale bar is 2.25 $\mu$m. (C) A graph representing the propulsion velocity dependence on the concentration of hydrogen peroxide. Red and orange horizontal bars are respectively the propulsion velocities glazed and dipped tori (with 40 nm Ni and 10 nm Pt) on a gold substrate at 0$\%$ H$_2$O$_2$. Green and yellow squares are respectively the propulsion velocities glazed and dipped tori (with 40 nm Ni and 10 nm Pt) on a gold substrate at 5$\%$ H$_2$O$_2$. Dark green diamond is the propusion velocity of glazed tori (with 40 nm Ni and 10 nm Pt)on a glass substrate at at 10$\%$ H$_2$O$_2$; and comparitively,teal diamond is the propulsion velocity of dipped tori (with 40 nm Ni and 10 nm Pt) on a gold substrate at 10$\%$ H$_2$O$_2$. Purple and dark blue triangles are respectively the propulsion velocities glazed (with 40 nm Ni and 10 nm Pt) and dipped tori (with 10 nm Ni and 40 nm Pt) on a gold substrate at 15$\%$ H$_2$O$_2$. Magenta, lime, and ice blue  circles are the propulsion velocity of 3 $\mu$m diamater glazed tori, 7 $\mu$m glazed tori, and 7 $\mu$m dipped tori at 30$\%$ H$_2$O$_2$. The three tori have a coating of 40 nm Ni and 10 nm Pt. A full legend can be found in the supplementary information.
	}  
	\label{fig3}
\end{figure}

\begin{figure}
	\includegraphics[width=\linewidth]{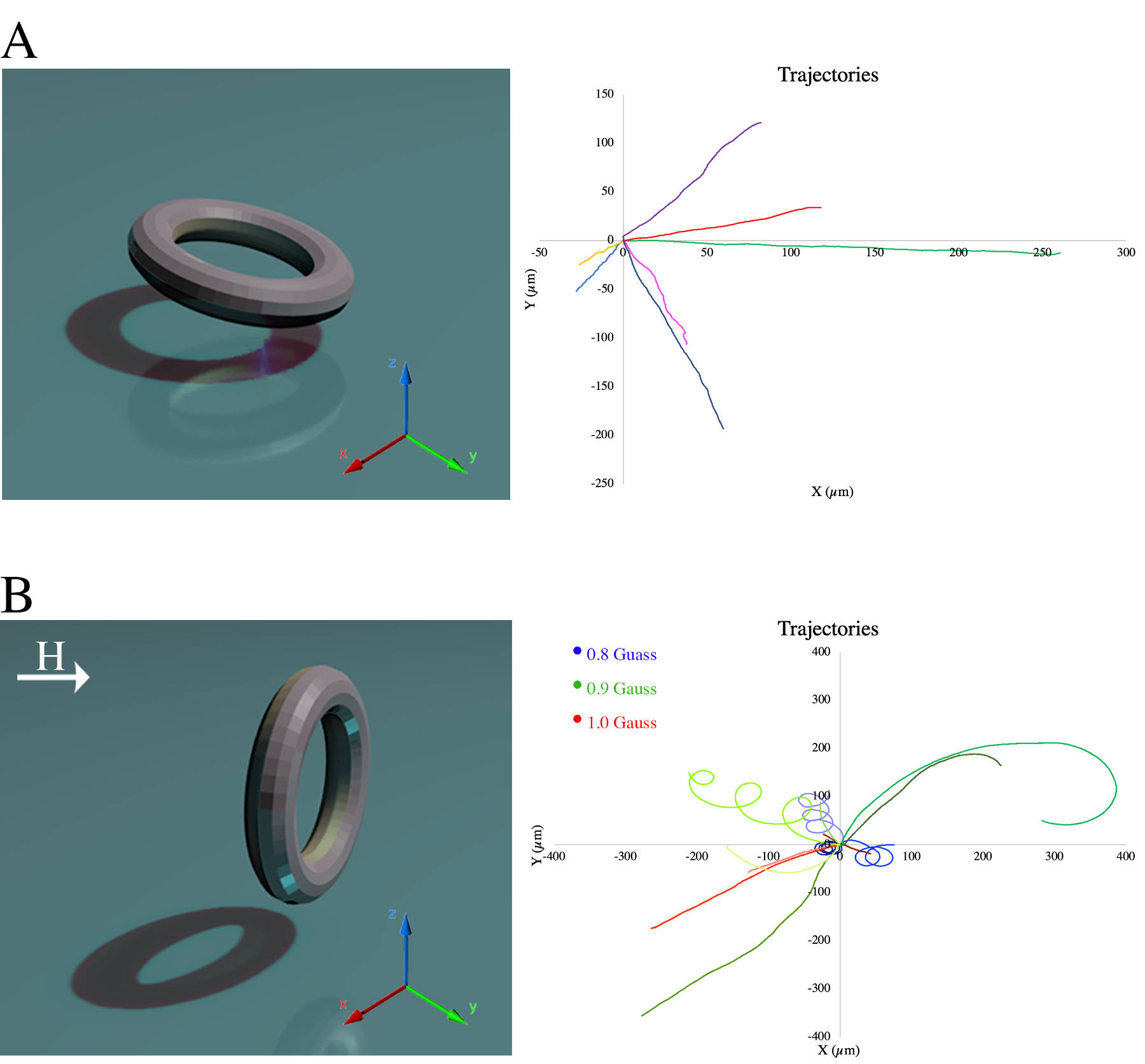}
	\caption{{\bf Swimming trajectories} (A) Extracted and re-centered trajectories for linearly translating dipped and glazed Janus tori.  (B) Trajectories of vertically-oriented tori in an increasing magnetic field. The trajectories change continuously from cyclodial to linear. Blue trajectories are for 0.8 G; green trajectories are for 0.9 G; and red trajectories are for 1.0 G.} 
	\label{fig9}
\end{figure}

\section{Experimental Set Up}
A colloidal dispersion was dispensed onto a gold-coated slide. The gold coating prevented microparticles from adhering irreversibly to the glass surface. Then 5-30\% hydrogen peroxide was introduced to the system to activate the catalytic particles. We protected the system from   evaporation by sealing the droplet with a cover slip.  Then the active particles were observed on an inverted Olympus Microscope IX83 at 10-100$\times$ magnification--oil immersion was specifically used for 60$\times$ and 100$\times$  magnification--and recorded at 11-56 frames per second with an Allied Vision ProSilica GT GIGE camera.

\section{Experimental Results} 
The microtori are 3D printed via two-photon lithography on a Nanoscribe Photonic Professional GT system.  After developing, we functionalize one face of the tori using electron beam metal evaporation; first depositing a thin binding nickel layer and then a catalytically active thin film of platinum. We investigate two different surface functionalizations: top "glazed'' and sideways "dipped'' Janus tori (Fig.\ref{fig3}A). See supplemental material for additional physical characterization of ``glazed'' and ``dipped'' Janus tori.

We extract the printed particles via pipetting and then drop cast the colloidal solution onto a gold-coated glass slide. In water, the tori exhibit pure Brownian diffusion. However, in the presence of hydrogen peroxide,  the 7$\mu$m glazed and dipped tori consume the fuel and directionally translate across the surface (Fig. \ref{fig9}A).  The slight decrease in velocity from glazed to dipped is the result of a decrease in the catalytic surface area on the torus. The propulsion velocity of the tori scales linearly with the concentration (v/v $\%$) of hydrogen peroxide up to 10 $\%$ and then saturates from 10-30 $\%$ (Fig. \ref{fig3}B) . In 30 $\%$ hydrogen peroxide, the glazed and dipped tori swim with the speed $\nu$ $\sim$ 7  $\mu$m s$^{-1}$ (Fig. \ref{fig3}B). 

Additionally, the deposition of a magnetic nickel binding layer (between the polymer and platinum) allows us to alter the orientation of the swimming tori with a magnetic field varying between 0.8 - 1.0 G. In the presence of a magnetic field, the tori reorient their axes to lie perpendicular ($\phi$ $\sim$ 90$^{\circ}$) to the substrate. For a weaker magnetic field $\sim$ 0.8 G, their trajectories are cycloidal (Fig. \ref{fig9}B). As the strength of the field increases to $\sim$ 1.0 G, vertically-oriented tori the rapidly alter their trajectories to be linear (Fig. \ref{fig9}B). We observe swimming trajectories in all directions, ruling out the migration in the magnetic field gradient. (See supplemental material for the estimation of the drift velocity.)

Furthermore, the vertically-oriented ($\phi$ $\sim$ 90$^{\circ}$) glazed tori swim significantly faster ($\nu$ $\sim$16.36 $\pm$ 2.49 $\mu$m s$^{-1}$) when compared to their horizontal orientation, see Fig. \ref{fig3}D. The increased velocity may be understood by considering both the increased drag present near a surface, and the direction of the propulsive force exerted by the torus; for perpendicular tori, all the propulsive force is used in linear translation, whereas for the gliding tori, a significant fraction of the force is used to hover over the boundary while only a small part on the order $\sin(15^{\circ}) = 0.26 \approx 1/4$ is used for linear translation.

\begin{figure}
	\includegraphics[width=\linewidth]{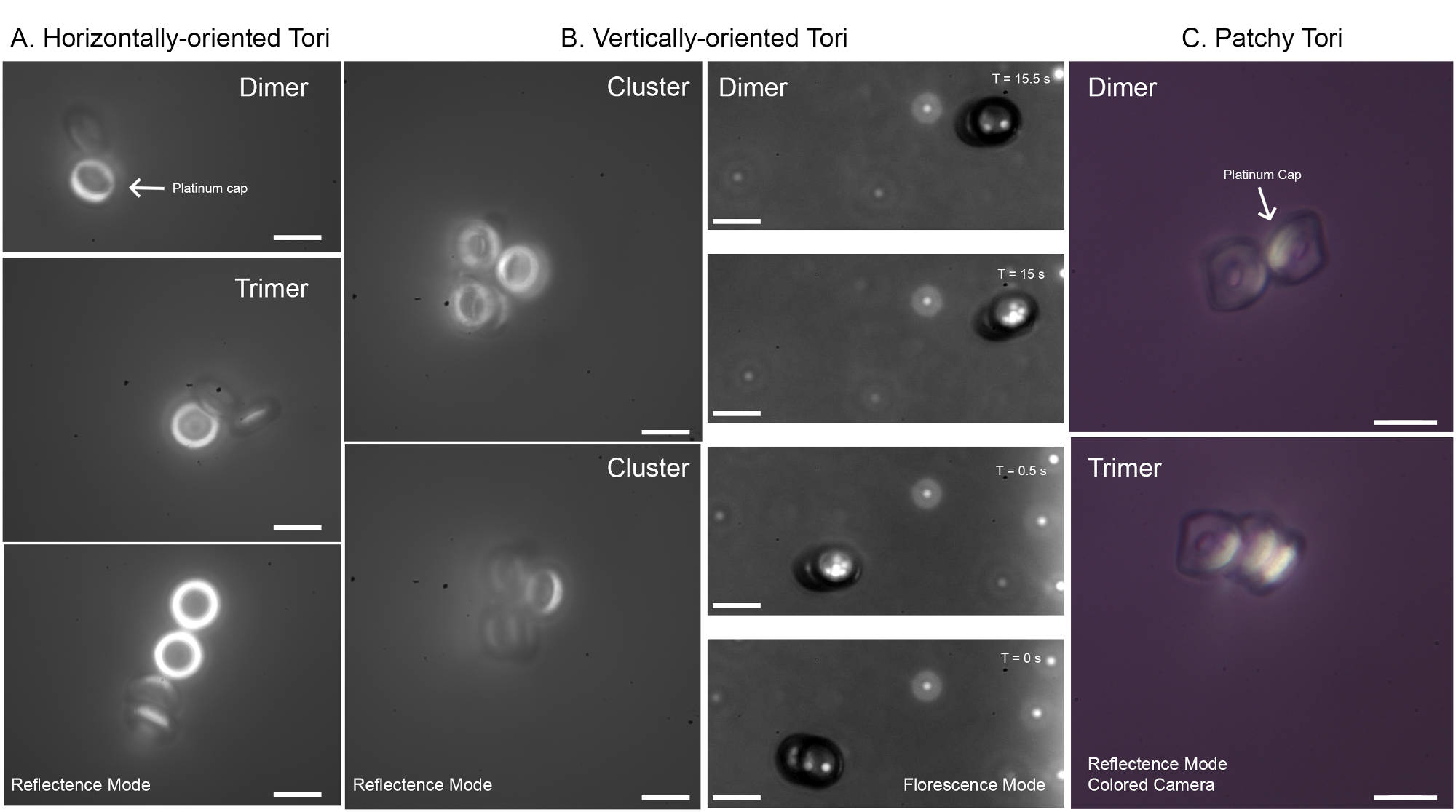}
	\caption{{\bf Collective states}  (A-B) Images of horizontally-oriented and vertically-oriented Janus tori clusters. Horizontally-oriented Janus particles form dynamic, unstable clusters. The vertically-oriented dimers swim in 3 dimensions. (C) An image of clustering, active patchy (dipped)  tori. The polymer is transparent,  and the metallic patches are reflected brightly in white. The dipped  tori attach to nearby particles at the catalytic cap. All scale bars are 6.5 $\mu$m.}
	\label{fig8} 
\end{figure}

We also observe formation of dynamic collective states of many active tori when in close proximity.  Similar to bimetallic nanorods and other Janus particles \cite{ginot2018aggregation}, Janus (glazed) tori form stable dimers, and the inclusion of more Janus tori leads to the formation of dynamic and unstable clusters (Fig. \ref{fig8}A,C). Trimers, tetramers, and similarly larger clusters spontaneously change configurations, ejected individual particles, and demonstrate quasi 2-D rotational and translational motion. For a complete study, we also investigate the behavior of vertically-oriented Janus tori. We find that Janus tori on overlapping trajectories collide, slide across each other's surfaces, and form stable tumbling dimers, Fig. \ref{fig8}B. Interestingly, stabilized vertical-dimers show a combined rotational and translation motion in 3D. Similar to their horizontally-oriented counterparts, large clusters of the vertical Janus tori show dynamic states with different conformations. 

We also analyze swimming patchy (dipped) tori. Tori with a thick nickel layer (40 nm) form stable dimers, trimers, and larger clusters. The dipped tori attach to other particles at the catalytic cap, a result arising from the strong magnetic dipoles, Fig. \ref{fig8}C. By contrast, in glazed tori with a thin nickel layer (10 nm), we observe active particles largely avoiding agglomerations. Our observation of the lack of clustering is similar to Ref. \cite{nourhani2017engineering}, who used a softly magnetic binding layer to induce contactless particle-particle interactions due to repulsion of parallel magnetic dipoles.

\section{Discussion} 

\subsection {Propulsion and Fluid Transport Mechanism}
We identify the dominant self-propulsion mechanisms that convertes the chemical energy in the surrounding hydrogen peroxide into linear and cyclodial swimming motions. First, our microscopic tori retain a slight negative charge arising from the polymer and repel to an equilibrium distance away from the negatively-charged substrate. This electrostatic repulsion prevents the particles from irreversibly adhering to the substrate.

Furthermore, we assum from previous Janus and patchy particle studies that the microscopic tori are self-electrophoretic.\cite{brown2014ionic, Ebbens2013} We test the hypothesis for a dominant self-electrophoretic propulsion mechansism by first including approximately 10 nM of salt (sodium chloride) into the hydrogen peroxide solution. The addition of salt inhibits self-electrophoretic swimming, and thus particles only move by diffusiophoresis. When we perform this experiment, we observe Brownian-like diffusive motion of the previously swimming microscopic tori and thus confirm our hypothesis that the dominant propulsion mechanism is self-electrophoresis.

In addition to the  self-electrophoretic propulsion mechanism, we investigat diffusion or advection driven fluid transport around the swimming tori. We use the dimensionless solute P\'eclet number, ${\rm Pe} =L u/D$, to relate the importance of advection to diffusion. Here $L$ is  the characteristic length, $u$ is  the fluid flow velocity, and $D$ is mass diffusion coefficient of the solute. Here, the characteristic length $L\approx 3-7$ $\mu$m is the diameter of the tori, the flow velocity $u\approx 10-20$ $\mu$m s$^{-1}$ is the propulsion velocity of the tori, and $D\approx 10^3$ $\mu$m$^2$ s$^{-1 }$ is the diffusivity of hydrogen peroxide in water. The estimate provides ${\rm Pe} \approx 0.1-0.2 $. The results indicate that the diffusion dominates the fluid advection  around the tori \cite{michelin2014phoretic}.

\subsection {The Onset  of Linear Motion}
Most synthetic swimmers display random walks unless their rotations are quenched  \cite{das2015boundaries, simmchen2016topographical} or their pathways are directed with external forces. However, we observe linear trajectories in swimming microscopic tori. Activating the microtori with hydrogen peroxide initiates fluid flows through the inner radius, i.e. donut hole, and outer radius that result in the particles hovering above the surface at a height $h$ $\sim$ 0.5-1 $\mu$m. We hypothesize that the hovering "parallel'' orientation is intrinsically unstable.  Any small fluctuations in the environment, e.g. thermal noise, leads to a small perturbation of the hovering parallel orientation of the tori. The small perturbation in the orientation, combined charge redistribution across the tori, and generation of electro-osmotic flow creates an instability surrounding the tori. The  instability leads to a spontaneous symmetry breaking \cite{snezhko2011magnetic, kaiser2017flocking} across the tori during which the tori tilt to a stable vertical angle. The exact vertical angle is determined by the electrostatic potentials between the charge substrate and the anode-cathode across the catalytically-active tori. The stabilized vertical angle, combined a lack of  rotational diffusion, gives rise to linear and persistent trajectories.

\subsection {Origin of the Cyclodial Trajectories}
We attribute cycloidal trajectories of vertically swimming tori to a misplacement of the ``center of propulsion'' and the 		``center of drag'' due to fabrication imperfections. 
In the presence of a permanent magnetic field $H$, the equations for the torus' center of mass (COM) are written in the dimensionless form: 
\begin{eqnarray}
\dot x &=&  V_0 \cos(\phi) \\
\dot y &=& V_0 \sin(\phi) \\ 
\dot \phi  &=&  \omega  +  H  \sin(\phi) t
\label{eq1} 
\end{eqnarray}
Here  $x,y$ represent the coordinates from the COM; $\phi$ is the in-plane angle; $V_0$ is  the self-propulsion speed; and $\omega$ is the dimensionless rotation frequency due to torque that arose from the misalignment of the drag and propulsion forces, $H \sin(\phi)$ represents the magnetic torque exerted from the field onto the torus. We assum for definiteness that the torus magnetic moment is oriented parallel to the torus axis of symmetry. Equation (\ref{eq1}) describes two regimes of motion. In the first regime where $H > \omega$, the angle $\phi$ approaches a stationary angle $\phi_{\rm s}$ that is defined in terms of $H,\omega$:  $\sin(\phi_{\rm s}) =\omega/H$. As a result, the COM moves in straight lines, as observed in experiments.
In the second regime where $H < \omega$, the trajectories become cycloids because of the juxtapoised translational and rotational motion. Then the average COM drift velocity $V_y = \langle \dot y  \rangle$ is obtained analytically for $0 < H < \omega$ using  Eq. (\ref{eq1}): $V_y = \langle \dot y  \rangle = \frac{1}{T} \int_0^T \dot y  dt$.

We use the period $T$  from Eq. (\ref{eq1}) and also assume from symmetry  $\langle \dot x \rangle=0$. The we solve analytically last  Eq. (\ref{eq1}) and find the period $T= 2 \pi/\sqrt{\omega^2-H^2}$. The period $T$ diverges for $H\to \omega$ and approaches $2 \pi/\omega$ for $H\to0$. For average velocity $V_y$ one obtaines:
\begin{equation}
V_y = \langle \dot y  \rangle = \frac{1}{T} \int_0^T \dot y  dt= \frac{1}{T} \int_0^{2 \pi}  \frac{ d y}{d \phi} d \phi =V_0  \frac{\sqrt{\omega^2-H^2}-\omega}{H}
\label{eq4}
\end{equation} 
Thus, from Eq. (\ref{eq4}), we find the averaged drift velocity was $V_y \to 0$ for $H \to 0$ and $V_y \to -V_0 $ for $H \to \omega$. This result indicate that the average drift velocity is zero when there was no magnetic field present (due to circular motion), and when a magnetic field was applied, the drift velocity gradually approaches the self-propulsion velocity with an increase of the magnetic field. For the values of magnetic fields $H > \omega$, a cycloidal motion transforms into a straight line. Furthermore, the torus drifts perpendicular to the direction of the magnetic field. In the general case, when the magnetic moment and torus axis of symmetry are not parallel, the expression for the mean drift velocity Eq. (\ref{eq4}) does not change; however the drift direction depends on the relative magnetic moment orientation.

\begin{figure*}
	\includegraphics[width=0.8\linewidth]{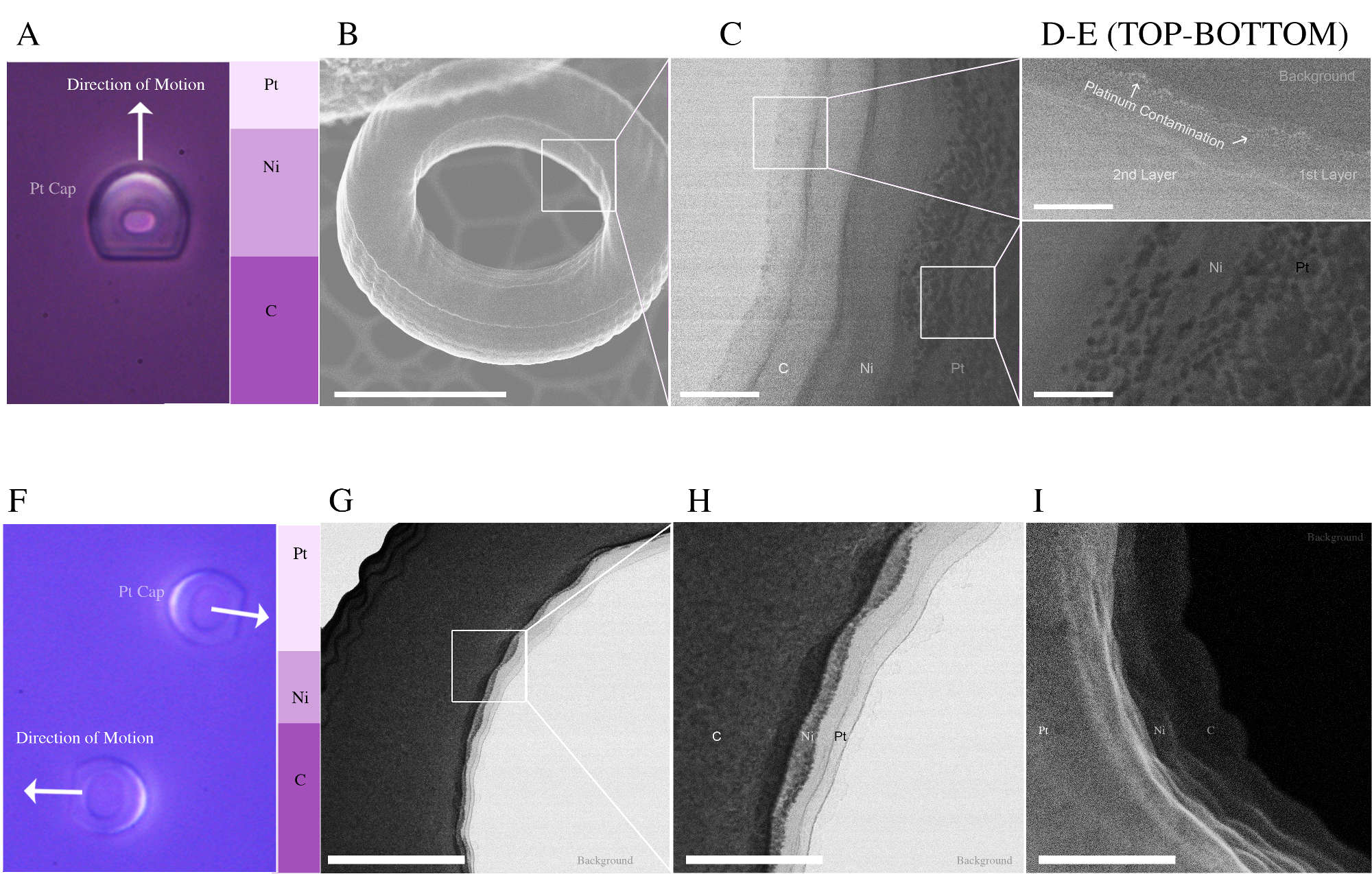}
	\caption{{\bf High-resolution TEM microgramms}(A) An image of the swimming orientation for tori with 40 nm nickel and 10 nm platinum. The Janus tori rest with their polymer structure near the substrate, and the patchy tori swim platinum-leading (B) A secondary electron HRTEM image focused on the cross section of a single layer in the 3D printed structure.  The scale bar is 3.25 $\mu$m. (C) A secondary electron HRTEM image of the different materials deposited on separate layers of the tori. The nickel forms a smooth structure while the platinum nucleates into droplets on the binding layer. The scale bar is 300 nm. (D) A bright-field image of platinum contamination (in white) on the first polymer layer in the tori. The platinum contamination is responsible for an enhanced conductivity on the surface of the polymer. The scale bar is 200 nm. (E) A dark-field image of the platinum droplets on the uniform nickel binding layer. The platinum formed droplets $\sim$ 20 nm in diameter. The scale bar is 100 nm. (F) An image of the swimming orientation for tori with 10 nm nickel and 40 nm platinum. The Janus tori rest with their metal cap near the substrate, and the patchy tori swim polymer-leading. (G) Cross-sectioned, dark-field image of a tori coated in 10 in nickel and 40 nm platinum. We see a clear separation between the carbon polymer, nickel binding layer, and catalytic platinum layer. The scale bar is 700 nm (H) A close-up, dark field image at the interfaces between platinum and and nickel. We observe a smooth platinum layer approximately 200 nm thick. (I) Secondary-electron imaging (on HRTEM) of the smooth platinum layer. The platinum layer covered almost entirely the nickel binding layer. The scale bar is 200 nm.}
	\label{fig7} 
\end{figure*}

\subsection {Programmable Swimming Orientation} Traditionally, most self-electrophoretic Traditionally, most self-electrophoretic Janus micromotors, namely microspheres, swim with the polymer-leading. \cite{das2015boundaries, simmchen2016topographical} Herein, we have observe two preferred swimming orientations, i.e. platinum or polymer face leading. We have find through High-Resolution Transmission Electron Microscopy (HRTEM) defects in the atomic interfaces among the polymer, nickel, and platinum that mainly contribute to the preferred swimming orientation (Fig. \ref{fig7}A, F). Our findings indicate that we can program preferred swimming orientation by purposefully including atomic scale defects in the thin films.

For tori that have $\sim$ 10 nm platinum layer deposited, we find a distinct lack of an anticipated uniform thin film. Instead, we observe $\sim$ 20 nm diameter platinum droplet features spattered across the nickel thin film (Fig. \ref{fig7}C-E). Gaps between the platinum clustering allow the exposed nickel surface to rapidly oxidize when exposed to atmosphere or hydrogen peroxide (Fig. \ref{fig7}E). We conclud that the presence of surface oxidized nickel alters the catalytic properties and electron mobility during self-electrophoresis \cite{paulus2002oxygen, kitajima1978formation, scarr1969mechanism, muvcka1985decomposition} through 180$^{\circ}$ flipping the positions of the anode-cathode in the tori.

In contrast, for tori where a $\sim$40 nm platinum layer is deposited, we observe the formation of a thin platinum film with three distinct grain boundaries (Fig. \ref{fig7}G-I). The platinum thin film completely covers the binding nickel layer; leaving nothing to be oxidized at atmospheres or in solution. Hence, tori with a uniform platinum film covering the nickel swim preferentially polymer side forward, as we observe experimentally and previously reported in other Janus swimmers.\cite{das2015boundaries, simmchen2016topographical}

\subsection {Controlled Collodial Crystal Formations} In close proximity and in moderately dense solutions, Janus particles form colloidal crystals \cite{palacci2013living, gangwal2008dielectrophoretic, yan2015rotating}. We demonstrate precise control over the microtori to form disordered aggregates or colloidal crystalline structures by utilizing the magnetic field to control their individual swimming behaviors. In the absence of a magnetic field, the microscopic tori translate across the substrate and collide with neighbors to form stable dimers. The dimers evolve into trimers, tetrameters, and larger structures (\ref{fig8}A). Similar to previous Janus colloidal crystals studies, we also observe dynamic interactions arising in trimers and larger structures due to the coupled electro- and hydrodynamics and self-generated chemical gradients.

With a magnetic field, we enforce the tori's orientation to be perpendicular to the substrate and consequently also alter the swimming behaviors from linear to cyclodial. The change in microtori orientation and behavior leads to the formation of collodial crystalline structures with rich dynamic behaviors. In particular, vertically-oriented microtori form dimers with 3-dimensional helical swimming patterns (Fig. \ref{fig8}B). We form ordered, crystalline structures with the addition of more microtori. 

\begin{figure}
	\begin{center}
		\includegraphics[width=0.65\linewidth]{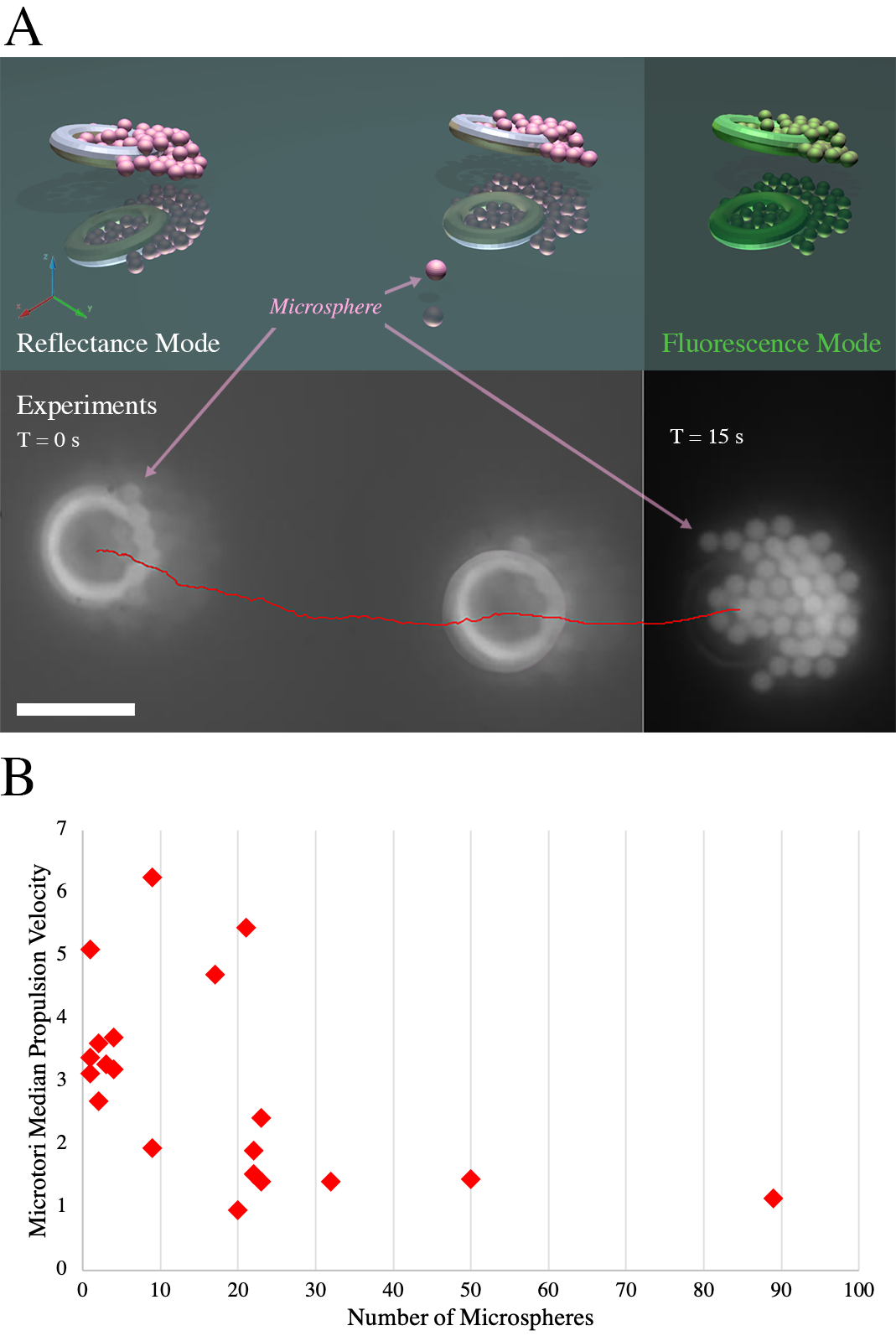}
		\caption{{\bf Passive cargo transport by swimming tori} (A) A time lapse sequence of a Janus donuts accumulating 1 $\mu$m sulfated polystyrene microspheres along the anode. The microspheres formed a hexagonal closed pack scheme around the inner and outer radii. The scale bar is 6.5 $\mu$m. (B) The propulsion velocity of the swimming tori versus number of attached microspheres.}
		\label{fig5} 
	\end{center}
\end{figure}

\subsection {Passive Cargo Manipulation.}
We also investigate the abilities of the swimming microscopic tori to pick-up and deliver cargo to specific sites. We use positively and negatively-charged microspheres as passive cargo. More specifically, we choose charged cargo to test the abilities of the tori to sort cargo based on charge.  We observe the tori sorting cargo based on charge (Fig. \ref{fig5} A). The tori preferentially collect negatively-charged 1 $\mu$m polystyrene microspheres along the leading anode.  Conversely, the tori accumulate $\mu$m positively charged latex spheres along the cathode. Notably, the tori reach a critical accumulation of microspheres and then arrange the microspheres into a ``halo'' around the outer circumference. At this critical cargo concentration, the propulsion velocity steeply drops. However, for tori that continue to swim past this critical cargo concentration, we find that their velocities are  then largely independent of the cargo concentration, Fig. \ref{fig5} B. Finally, we utilize a strong external magnetic field to precisely direct tori carrying cargo. We also apply a magnetic pulse to release the cargo on-demand.

\subsection {Active Cargo Manipulation.} 
\begin{figure}
	\includegraphics[width=\linewidth]{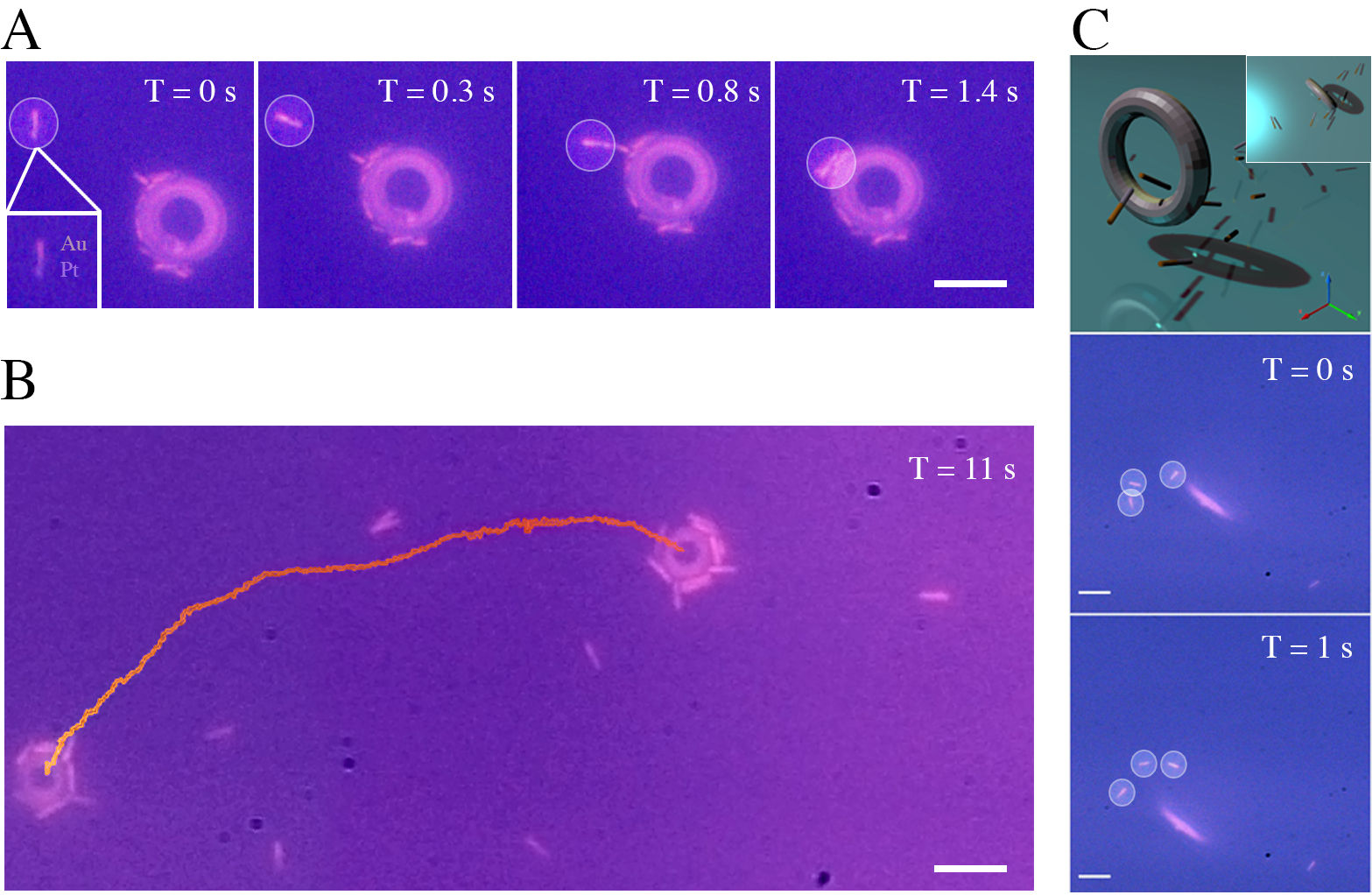}
	\caption{{\bf Active cargo transport by swimming tori} (A) A timelapse showing the hydro- and electrodynamic attachment of a swimming bimetallic nanorod to a microtori. The scale bar is 7 $\mu$m. (B) A timelapse showing a 3 $\mu$m diameter microtori transporting numerous bimetallic nanorods. The scale bar is 2.5 $\mu$m. (C) A cartoon showing the vertical orientation of the micoscopic tori relative to the bimetallic nanorods near the surface. The bimetallic nanorods align along the self-generated fluid streamlines of the microscopic tori when nearby. The scale bar is 2.5 $\mu$m.
	}
	\label{fig10} 
\end{figure}

For possible applications in living cell transport and sorting, we explore the complex task of manipulating and transporting active matter, i.e. other self-propelled microagents. We fabricate 2 $\mu$m long self-propelled, gold-platinum nanorods for their {\it E. coli}  biomimicry. \cite{Paxton2006} We observe two distinct modes of active cargo manipulation that depend upon the orientation of the tori. In the first mode, horizontally-oriented tori actively collect swimming nanorods around their outer circumference and transport the active agents, Fig. \ref{fig10} A-B. We switch to a second mode of manipulation by orienting the tori perpendicularly to the substrate. In this vertical orientation the tori manipulate the swimming behaviours of the bimetallic nanorods, Fig. \ref{fig10} C. More specifically, the active nanorods orient along the self-generated streamlines of the tori; either being pushed away from the tori along the outer radial streamlines or made to ``jump through a [toroid's] hoop." In addition, the tori can be continuously tuned between both modes by a magnetic field for a combination of active matter sorting and transport.

\subsection{Physical Description for Swimming Torus Dynamics.}
We model a single torus with an active surface (catalytic layer) oriented upwards and away from a substrate or plane boundary. We also vary the density $\rho$ of the torus; starting with $\rho_a=0$ representing a fluid density-matched (neutrally-buoyant) swimmer. Far from the substrate, the neutrally buoyant swimmer swims vertically upwards, with decreasing speed as the density increases. At a critical density $\rho_c$, the torus hover, and at higher densities the torus sediments.

\begin{figure}[t]
	\begin{center}
		\includegraphics{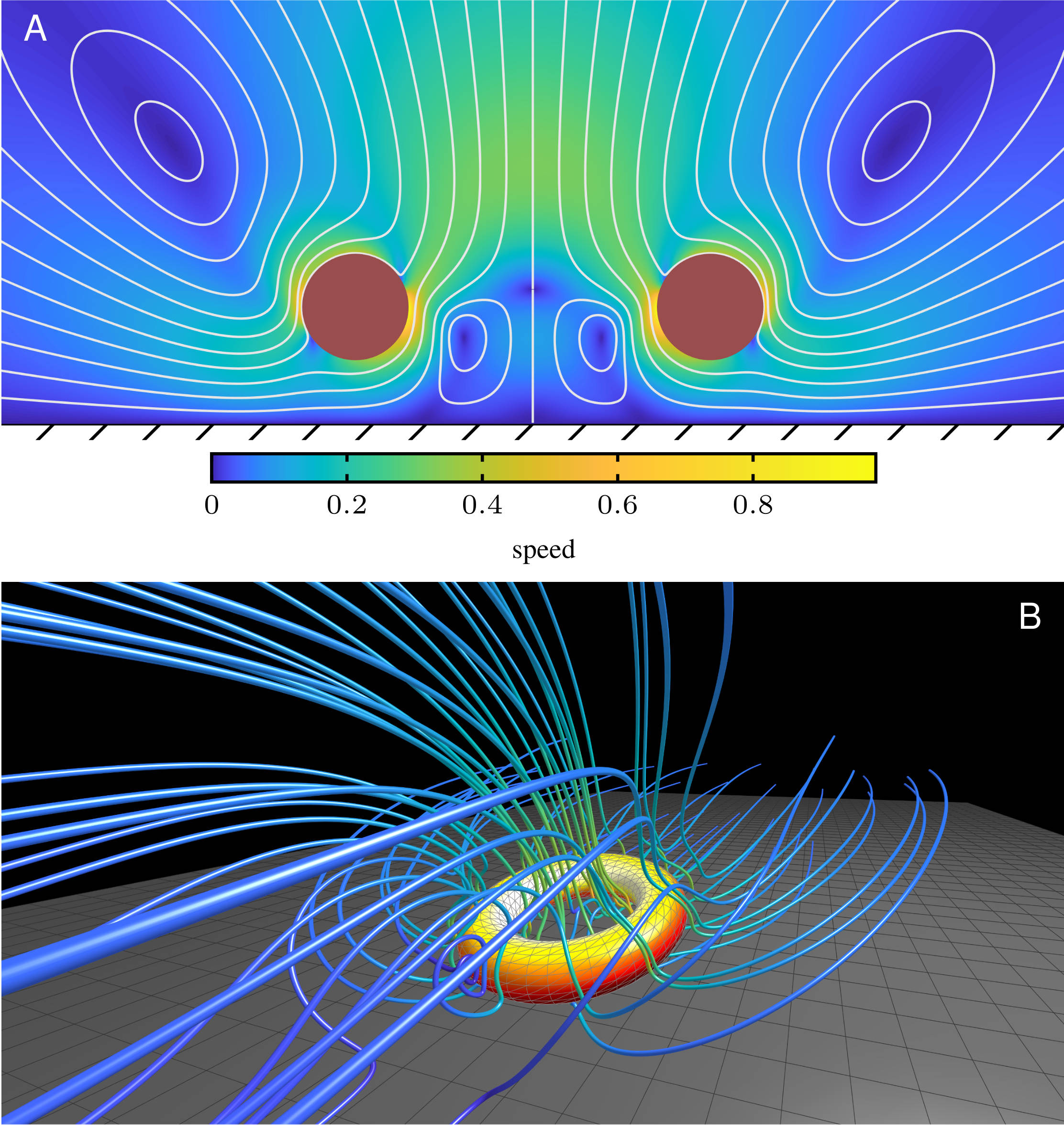}
	\end{center}
	\caption{{\bf Flow streamlines generated by a patchy (glazed) torus} (A) Flow streamlines produced by an active, uncharged torus, hovering above a plane boundary, with nondimensionalized speed. (B) Flow streamlines and surface concentration of a steady gliding torus, at an angle of $12^\circ$, shown in the frame in which the torus is stationary.}
	\label{fig:hovering_streamlines}
\end{figure} 

\begin{figure}[t]
	\includegraphics{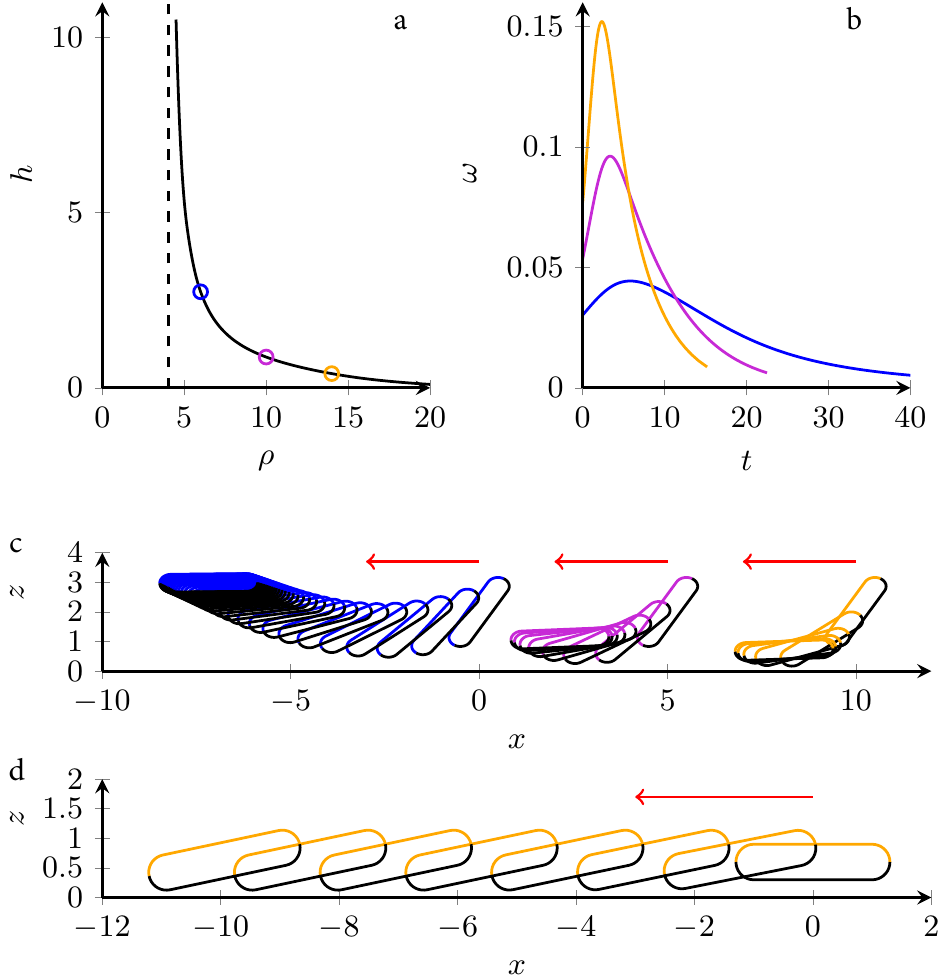}
	\caption{{\bf Simulated dynamics of tori over a plane boundary} (A) The height $h$ at which the base of a torus with unit activity and mobility hovers over the plane boundary as a function of its added density $\rho$ (where $\rho=0$ is neutrally buoyant). The asymptote $\rho = 4.33$ is the density at which such a torus in free space neither sinks nor swims. (B) The relaxation to a stable horizontal hovering state of uncharged tori, showing decaying angular velocity $\omega$ with time. 
		(C) Swimmer trajectories with equal intervals of uncharged tori.  
		(D) Transition from horizontal to steady gliding state (at a 12$^\circ$  angle) of a charged torus. Red arrows indicate progression of time. Here  the major and minor radii of the tori $a_1=1, a_2 = 0.3$. }
	\label{fig:results}
\end{figure}

The behavior of a sedimenting self-propelled torus changes significantly as it approaches the plane boundary at $z=0$; the torus gains a propulsive boost by ``pushing off'' the substrate with the flow it generates. For densities above the critical density, the torus approaches a critical height $z_0(\rho),\ \rho > \rho_c,$ where the swimmer begins to hover above the substrate via self-propulsion (autophoretic slip flows) (Fig.~\ref{fig:hovering_streamlines}A). We measure the heights above the plane boundary $h$ from the bottom of the torus as a function of density $\rho$. As the  density of the torus increases we observe a decrease in the hovering heights (Fig.~\ref{fig:results}A). 

We also consider the effects of electric charge on a hovering torus. In the absence of charge, a vertical position of a torus hovering via autophoretic slip flows above a plane boundary is stable. We perturbe the active torus to a non-zero angle and observe no appreciable changes in the slip velocity on the portion of the torus closest to the boundary. However, the portion of the torus closest to the boundary exerts more thrust than the other side and results in a restorative torque. We fully capture the relaxation to a stable configuration for different torus densities (Fig.~\ref{fig:results} B-C). 

For the second case we consider a hovering torus near a boundary with charge. We include a dipolar charge distribution across the active surface; we also set a positive charge on the boundary due to surface charge absorption. The torque tilts the swimmer away from the parallel to perpendicular state (assuming there were no collisions with the plane boundary). Combined with the presence of a solute fuel-driving (diffusiophoretic) surface slip flow, the torus glides at a stable angle; the exact stable angle is determined from the competing electrostatic attraction / repulsion forces and restorative torque from the slip flow. We capture the dynamics of an initially horizontally-oriented torus that quickly reaches a steady angle of 12$^\circ$  as it glides across the surface (Fig.~\ref{fig:results}D).  The flow streamlines and surface solute concentration are shown for the example in Fig.~\ref{fig:hovering_streamlines}B. For simplicity, we did not use any short range repulsive potential; however, depending on the relative charge, short range repulsion may be what prevents the leading edge from adhering to the lower boundary.

\section{Conclusions}

We utilize 2-photon lithography to 3D print and program by shape autonomous, multi-responsive behaviors in artificial microswimmers. We fabricate chemically-powered microscopic tori with nanoscale features; these tori have either half-coated (glazed or Janus) or patchy (dipped) platinum catalytic layers. In the presence of hydrogen peroxide, the tori instantaneously begin to hover above the surface because of the self-electrophoretic propulsion. Due to the charge instability, i.e. moving surface charges arising from self-electrophoresis, the tori spontaneously break symmetry and tilt to a stable angle. The tori then glide across the surface; eventually organizing into dynamic clusters that swim in three dimensions. 

Furthermore, we program two modes, i.e. linear or cycloidial swimming behaviors, into the tori to perform complex tasks. For a simple task, the tori scavenge and sort passive and charged latex microspheres in solution; the tori organize the tracers into a hexagonal closed packing. At the same time, the swimming tori are also capable of sorting tracer particles across their anode-cathode. We demonstrate that the cargo can be delivered to a desired location and released on demand.  From these results, it is conceivable to utilize our motile and programmable swimmers to clean freshwater of pollutants, such as microplastic particles. 

We also task the microtori with a more complex objective--the manipulation of  active matter. In the first mode the tori swim linearly and almost parallel to the substrate; the swimmers accumulate and transport very dense, self-propelled bimetallic nanorods. Switching to the second mode, the tori reorient perpendicular to the substrate and manipulate the orientation of nearby nanorods. The bimetallic nanorods align along the  self-generated fluid streamlines of the tori. The tori then either draw in through the center radii or push away nearby nanorods. To our knowledge, this is the first such manipulation of active matter by autonomous, artificial swimmers. 

The concepts and results presented here can be further extended for biological applications. In particular there are many options available for biocompatible catalysts, including tethered enzymes or light-driving propulsion mechanisms. The addition of more complex external fields and chemical gradients can be used to indirectly and directly guide the autonomous swimmers. The tori could then be directed to deliver living cargo, such as cells, to specific sites for cell therapy; or collectively organize the tori to direct their flow for cellular transport and sorting. 

\bibliographystyle{naturemag}
\bibliography{references}

\begin{addendum}
 	\item This research was supported by Penn State MRSEC DMR-1420620 grant (R.D.B.), NSF DMREF grant DMS-1628411 (A.S.), and NSF DMREF grant DMS-1735700 (ISA). Contributions from TDM-J were supported by EPSRC Grant No. EP/R041555/1. This project has also received funding from the European Research Council (ERC) under the European Union's Horizon 2020 research and innovation programme (grant agreement 682754 to E.L.). We also would like to acknowledge the Center for Nanoscale Science and their staff at the Pennsylvania State University for their contributions in fabrication. Additionally, financial support for characterization came from NSERC Canada. We would also like to thank the OCCAM-University of Toronto team where all the microscopy was performed. Lastly, we thank F.D.C. Isaac for their contributions in editing and discussions.
 	  
 	\item[Author contributions.] IAS and AS conceived the research. RB  fabricated, performed the experiments, and analyzed the behavior of swimming tori.  ADS and MJT characterized via electron microscopy the tori. TDMJ and EL performed the simulations for swimming tori. All authors discussed the results and wrote the paper. 
 	\item[Competing Interests.] The authors declare that they have no
 	competing financial interests.
 	\item[Correspondence.] Correspondence and requests for materials
 	should be addressed to I.S.A.~(email: isa12@psu.edu).
 \end{addendum}

\end{document}